\documentclass[11pt,prd,preprintnumbers,nofootinbib,superscriptaddress,notitlepage]{revtex4-1}

\usepackage{graphics}
\usepackage{bm}
\usepackage{cancel}
\usepackage{color}
\usepackage{xcolor}
\usepackage{multirow}
\usepackage{slashed}
\usepackage{array}
\usepackage{amssymb}
\usepackage{graphicx}%Í¼Æ¬ºê°ü
\usepackage{mathrsfs}% »¨Ìå×ÖÄ¸
\usepackage{diagbox}% ±í¸ñÍ·Ð±Ïß
\usepackage{amsmath}%×Ó¹«Ê½
\usepackage{float}%´ò¿ªÓÐÂ·¾¶ÎÄ¼þºê°ü
\usepackage{extarrows}% ¼ýÍ·¹¦ÄÜ
\graphicspath{{figures/},{pics/}}
\usepackage[colorlinks,citecolor=blue,anchorcolor=red,menucolor=red,linkcolor=red,filecolor=red,runcolor=red,urlcolor=red,frenchlinks=red]{hyperref}
\usepackage{subfigure}

\makeatletter
\newcommand{\thickhline}{\noalign {\ifnum 0=`}\fi \hrule height 1pt\futurelet \reserved@a \@xhline}
\newcolumntype{"}{@{\hskip\tabcolsep\vrule width 1pt\hskip\tabcolsep}}                             
\makeatother

\def\nn{\nonumber}

\def\pslash{\not{\hbox{\kern-4pt $p$}}}
\def\qslash{\not{\hbox{\kern-4pt $q$}}}
\def\lv{\not{\hbox{\kern-4pt $L$}}}
\def\lsim{\mathrel{\raise.3ex\hbox{$<$\kern-.75em\lower1ex\hbox{$\sim$}}}}
\def\gsim{\mathrel{\raise.3ex\hbox{$>$\kern-.75em\lower1ex\hbox{$\sim$}}}}
\def\ifmath#1{\relax\ifmmode #1\else $#1$\fi}

\begin{document}
\title{Search for heavy Majorana neutrinos at future lepton colliders }

\author{Peng-Cheng Lu}
\email{pclu@sdu.edu.cn}
\affiliation{School of Physics, Shandong University, Jinan, Shandong 250100, China}

\author{Zong-Guo Si}
\email{zgsi@sdu.edu.cn}
\affiliation{School of Physics, Shandong University, Jinan, Shandong 250100, China}

\author{Zhe Wang}
\email{wzhe@mail.sdu.edu.cn}
\affiliation{School of Physics, Shandong University, Jinan, Shandong 250100, China}
	
\author{Xing-Hua Yang}
\email{yangxinghua@sdut.edu.cn}
\affiliation{School of Physics and Optoelectronic Engineering, Shandong University of Technology, Zibo, Shandong 255000, China}	
	
\author{Xin-Yi Zhang}
\email{xinyizhang@mail.sdu.edu.cn}
\affiliation{School of Physics, Shandong University, Jinan, Shandong 250100, China}
	
%\date{\today}

\begin{abstract}
The nonzero neutrino mass can be a signal for new physics beyond the standard model. To explain the tiny neutrino mass, we can extend the standard model with right-handed Majorana neutrinos in a low-scale seesaw mechanism, while the CP violation effect can be induced due to the CP phase in the interference of heavy Majorana neutrinos. The existence of heavy Majorana neutrinos may lead to lepton number violation processes, which can be used as a probe to search for the signal of heavy Majorana neutrinos. In this paper, we focus on the CP violation effect related to two generations of heavy Majorana neutrinos for $15$ GeV $<m_N<$ $70$ GeV in the pair production of $W$ bosons and rare decays. It is valuable to investigate the Majorana neutrino production signals and the related CP violation effects in the W boson rare decays at future lepton colliders.                        
\end{abstract}


\maketitle

\section{INTRODUCTION}\label{sec1}

The neutrinos in the standard model (SM) are strictly massless due to the absence of right-handed chiral states and the requirement of $SU(2)_L$ gauge invariance and renormalizability. However, the neutrino oscillation experiments show that the neutrinos have minor non-zero masses, this insinuates the existence of new physics beyond the SM. In order to explain the tiny neutrino mass, we need to extend the SM. The mechanism with respect to the neutrino mass generation is attracting more and more attention. An interesting mechanism to generate the neutrino mass which introduces the heavy right-handed Majorana neutrinos ($\overline{N^c_R}$) is known as type-I seesaw mechanism~\cite{Minkowski:1977sc,Yanagida:1979ss,Gell-Mann:1979ss,Glashow:1979ss,Mohapatra:1979ia}. In this mechanism, the Dirac neutrino mass terms will be generated after spontaneous gauge symmetry breaking. In addtion, the Majorana neutrinos can also form Majorana mass terms $\overline{N^c_R}M_RN_R$. The introduction of heavy Majorana neutrinos may lead to the CP violation which gives a reasonable explanation to the asymmetry of matter and antimatter. The asymmetry of baryon number in our Universe can be measured by the baryon-to-photon density ration $\eta=n_B/n_\gamma$. According to an analysis of recent Planck measurement of the cosmic microwave background, the value of $\eta$ is about $\eta\simeq(6.12\pm0.03)\times10^{-10}$~\cite{Planck:2018vyg}. This result is too large compared to the result which SM expects. In the famous mechanism called leptogenesis, this asymmetry of baryon number can be produced by the lepton number asymmetry caused by the decay of Majorana neutrinos in the early Universe, the latter may turn into baryon number asymmetry through sphaleron processes \cite{Manton:1983nd,Klinkhamer:1984di} in this situation. The existence of the heavy Majorana neutrinos also leads to the violation of lepton number by two units $\Delta L=2$, such as the neutrinoless double-beta decay $(0\nu\beta\beta)$ \cite{Furry:1939qr,Elliott:2004hr}. In our study, we consider rare $W$ decay through Majorana neutrinos exchange which will produce the same-sign dilepton in the final state. The process of $W$ rare decay $W^-\rightarrow\ell_1^-N\rightarrow\ell_1^-\ell_2^-(q\bar{q}^\prime)^+$ has been studied in the literature(for a review, see e.g. Refs. \cite{Deppisch:2015qwa,Antusch:2016ejd,Cai:2017mow,Lu:2022pvw,Lu:2021vzj}). At the mass range of $M_N<M_W$ , the $\Delta L=2$ same-sign dilepton production signal
 has been studied in rare meson decays \cite{Atre:2005eb,Cvetic:2010rw,Wang:2014lda,Dong:2013raa,Cvetic:2016fbv}, tau lepton decays \cite{Gribanov:2001vv,Kobach:2014hea,Yuan:2017xdp} and even top quark decays \cite{Bar-Shalom:2006osy,Si:2008jd,Delepine:2012nea,Liu:2019qfa}. There is a nonzero CP asymmetry  induced by the difference between the rates of $W^+\rightarrow\ell_1^+\ell_2^+(\bar{q}q^\prime)^-$ and its CP-conjugate process $W^-\rightarrow\ell_1^-\ell_2^-(q\bar{q}^\prime)^+$. This difference arises from the significant interference of different heavy Majorana neutrinos. We can use the lepton number violation (LNV) process as a probe to study the heavy Majorana neutrinos. Theorists have done a great number  work on examining the prospects for observing CP asymmetry via LNV processes in which the CP asymmetry can be observed in the decays of mesons \cite{Cvetic:2015naa,Cvetic:2020lyh,Godbole:2020jqw,Zhang:2020hwj} and tau leptons \cite{Zamora-Saa:2016ito,Tapia:2019coy}. Recently, Ref. \cite{Najafi:2020dkp} studied the CP asymmetry in rare $W$ decays at LHC, but the CP violation effect produced in the situation they considered is influenced by the initial parton distribution functions in proton. In this work, we investigate the prospects for searching for heavy Majorana neutrinos in future lepton colliders like the Compact Linear Collider (CLIC)~\cite{Linssen:2012hp}, the International Linear Collider (ILC)~\cite{Behnke:2013xla} and the Muon Collider (MuC)~\cite{MuonCollider:2022cre} which provide a much cleaner environment. We analyse the process of $W^\pm$ pair production in $e^+e^-$ collison with ILC running at 190 GeV and $\mu^+\mu^-$ collison with MuC running at 190 GeV, where the $W^+$ and $W^-$ arise from $e^+e^-$ and $\mu^+\mu^-$ collision, and the influence of the parton distribution functions in proton on the CP violation can be cancelled. In principle, the number $n$ of right-handed heavy Majorana neutrinos introduced in type-I seesaw mechanism is a free parameter. We take $n\geq 2$ as two neutrino mass-squared differences between light neutrinos have been observed. Another reason is that the CP violation requires the interference of at least two different Majorana neutrinos. For convenience, we take only two heavy Majorana neutrinos $N$ and $N_b$ into consideration, the general case with more Majorana neutrinos can be analyzed in a similar way.

This paper is organized as follows. The model including heavy Majorana neutrinos is briefly reviewed in Section \ref{sec2}. Then we investigate the W pair production and their rare decays via Majorana neutrinos at future lepton colliders in Section \ref{sec3}. Finally, a short summary is given in Section \ref{sec4}.     

\section{BRIEF REVIEW OF THEORETICAL MODEL}\label{sec2}

An effective extension of the SM introduces three right-handed neutrinos, which are singlets under the SM gauge group. The CP violation stems from different Majorana neutrinos, for convenience, we only consider two generations of Majorana heavy neutrinos, which are denoted as $N$, $N_b$. We can write the Lagrangian of the model by \cite{Mekala:2022cmm}:
\begin{equation}
\label{1}
\mathcal{L}=\mathcal{L}_{SM}+\mathcal{L}_{N}+\mathcal{L}_{WN\ell}+\mathcal{L}_{ZN\nu}+\mathcal{L}_{HN\nu}
\end{equation}
where $\mathcal{L}_N$ is a sum of kinetic and mass terms for heavy neutrinos:
\begin{equation}
	\label{2}
	\mathcal{L}_N=\xi_\nu \left(\bar{N_k}i\partial\!\!\!/N_k-m_{N_k}\bar{N_k}N_k\right)  
\end{equation}
where $k=1,2$, and the overall factor $\xi_\nu=1$ for the Dirac neutrinos and $\xi_\nu=\frac{1}{2}$ for Majorana neutrinos. In our case $\xi_\nu=\frac{1}{2}$. The $\mathcal{L}_{WN\ell}$ corresponds to heavy neutrino interactions with a $W$ boson:  
\begin{equation}
	\label{3}
\mathcal{L}_{WNl}=-\frac{g}{\sqrt{2}}W_\mu^+\sum_{k=1}^{2}\sum_{l=e}^{\tau}\bar{N_k}R_{l k}^\ast\gamma^\mu P_L\ell^- + h.c.,
\end{equation}
The $\mathcal{L}_{ZN\nu}$ to interactions with a Z boson:
\begin{equation}
	\label{4}
	\mathcal{L}_{ZN\nu}=-\frac{g}{2\text{cos}\theta_{W}}Z_{\mu}\sum_{k=1}^{2}\sum_{l=e}^{\tau}\bar{N_k}R_{lk}^\ast\gamma^{\mu}P_L\nu_{l} + h.c,
\end{equation}
at last the $\mathcal{L}_{HN\nu}$ to interactions with a Higgs boson:
\begin{eqnarray}
	\label{5}
	\mathcal{L}_{HN\nu}=-\frac{gm_N}{2M_W}h\sum_{k=1}^{2}\sum_{l=e}^{\tau}\bar{N_k}R_{lk}^\ast  P_L\nu_l + h.c.	
\end{eqnarray}
In our process we only use the $\mathcal{L}_{WNl}$. Then we can write the weak charged-current interaction Lagrangian as 
\begin{align}
	\label{6}
	\mathcal{L}_{\rm cc}
	= -\frac{g}{\sqrt{2}} W^{+}_{\mu} \sum_{\ell=e}^{\tau} \sum_{m=1}^{3} V_{\ell m}^{\ast} \overline{\nu_{m}} \gamma^{\mu} P_{L} \ell
	- \frac{g}{\sqrt{2}} W^{+}_{\mu} \sum_{\ell=e}^{\tau} \sum_{k=1}^{2} R_{\ell k}^{\ast} \overline{N^{c}_{k}} \gamma^{\mu} P_{L} \ell + h.c. \; 
\end{align}
Here the $V_{\ell m}$ is neutrino mixing matrix and we can measure its parameters from the neutrino oscillation experiments. We neglect the contributions of the light Majorana neutrinos to the LNV processes since their small masses. The $R_{\ell k}$ are the mixing elements between heavy Majorana neutrinos and charged-leptons, we can parameterize them as~\cite{Xing:2007zj}
\begin{equation}
	\label{7}
	R_{\ell k}=\left|R_{\ell k}\right|e^{i\phi_{\ell k}} , ~~~ \ell= e, \mu, \tau, ~~~ k=1, 2 \; .
\end{equation}
Now we give the mixing relations between the neutrino flavor eigenstates and mass eigenstates as follow:\cite{Atre:2009rg} 
\begin{equation}
	\label{8}
	\nu_{\ell L} = \sum_{m=1}^{3} V_{\ell m} \nu_{m L} + \sum_{k=1}^{2} R_{\ell k} N^{c}_{kL} \; .
\end{equation}
Complex phases $\phi_{\ell k}$ give contribution to CP violation when there are interference from two different heavy Majorana neutrinos and can be determined in possible collider experiments. At present, the constraints on the mixing parameters between heavy Majorana neutrinos and leptons are studied in previous works \cite{CMS:2018iaf,Tastet:2021vwp}. According to the constraints on the mixing parameters, at the range of $15$ GeV $<m_N<70$ GeV, the parameters in the following range all satisfy with the constraints: 
\begin{eqnarray}
	\label{9}
	\left|R_{e i}\right|^2 \leq8\times10^{-6}   \; ,~~ \left|R_{\mu i}\right|^2 \leq1\times10^{-5},~~ \left|R_{\tau i}\right|^2 \leq1\times10^{-5}  \; , \; {\rm for} ~~ i = 1, 2 \; .
\end{eqnarray}
For convenience, we set these parameters as $\left|R_{e i}\right|^2 =8\times10^{-6}, \left|R_{\mu i}\right|^2 =1\times10^{-5}, \left|R_{\tau i}\right|^2 =1\times10^{-5}$, and we take the situation that two generations of heavy Majorana neutrinos are nearly degenerate where $\Delta m=m_{Nb}-m_N$. We also consider the result when $\Delta m=m_{Nb}-m_N=20$ GeV when we want to see the effects of $\Delta m$ in our work.

\section{$W^+W^-$ PAIR PRODUCTION AND SEARCH FOR HEAVY MAJORANA NEUTRINOS AT FUTURE LEPTON COLLIDERS}\label{sec3}
\begin{figure}[!h]
	\begin{center}
		\subfigure[]{\label{fig1a}
			\includegraphics[width=0.45\textwidth]{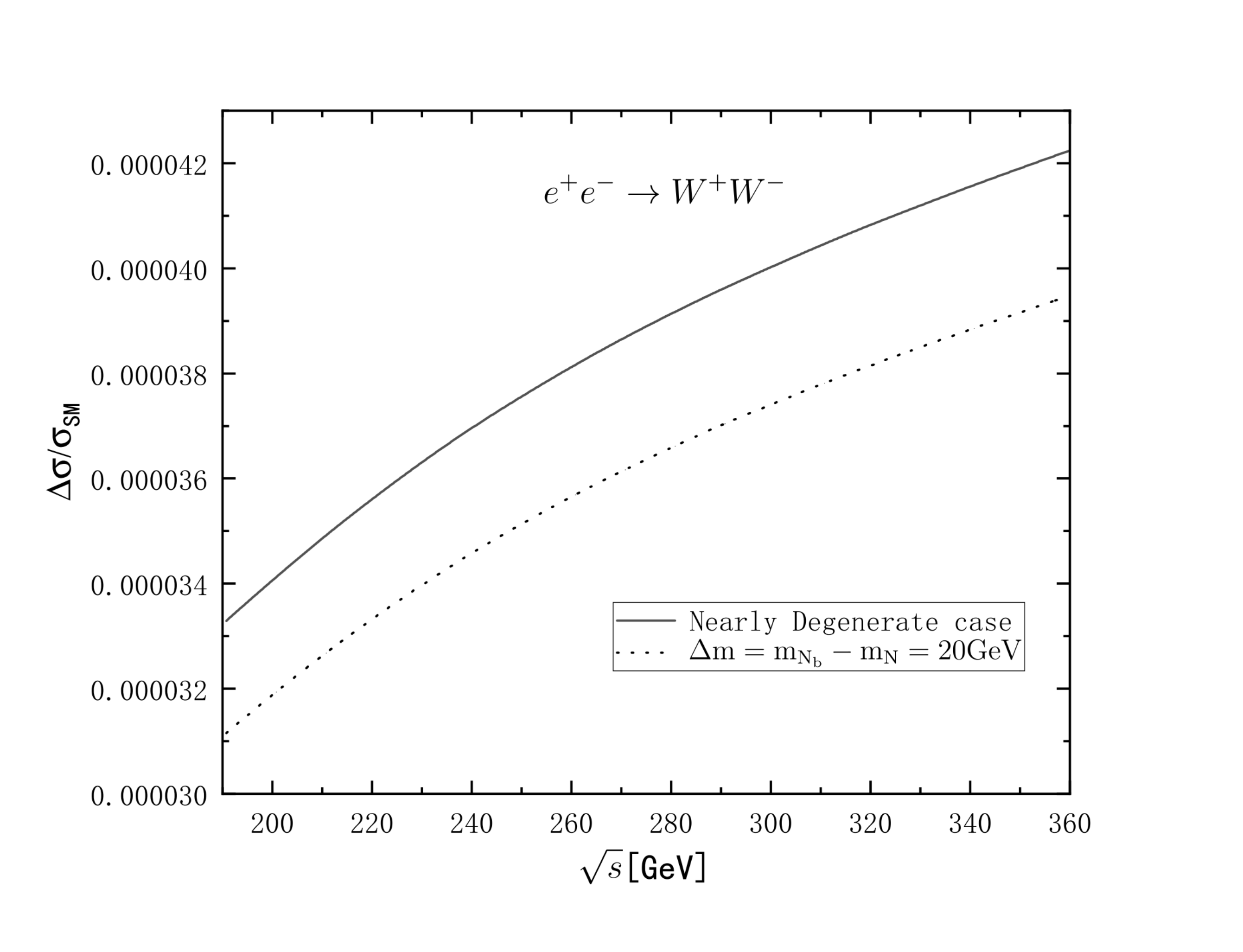} }
		\hspace{-0.5cm}~
		\subfigure[]{\label{fig1b}
			\includegraphics[width=0.45\textwidth]{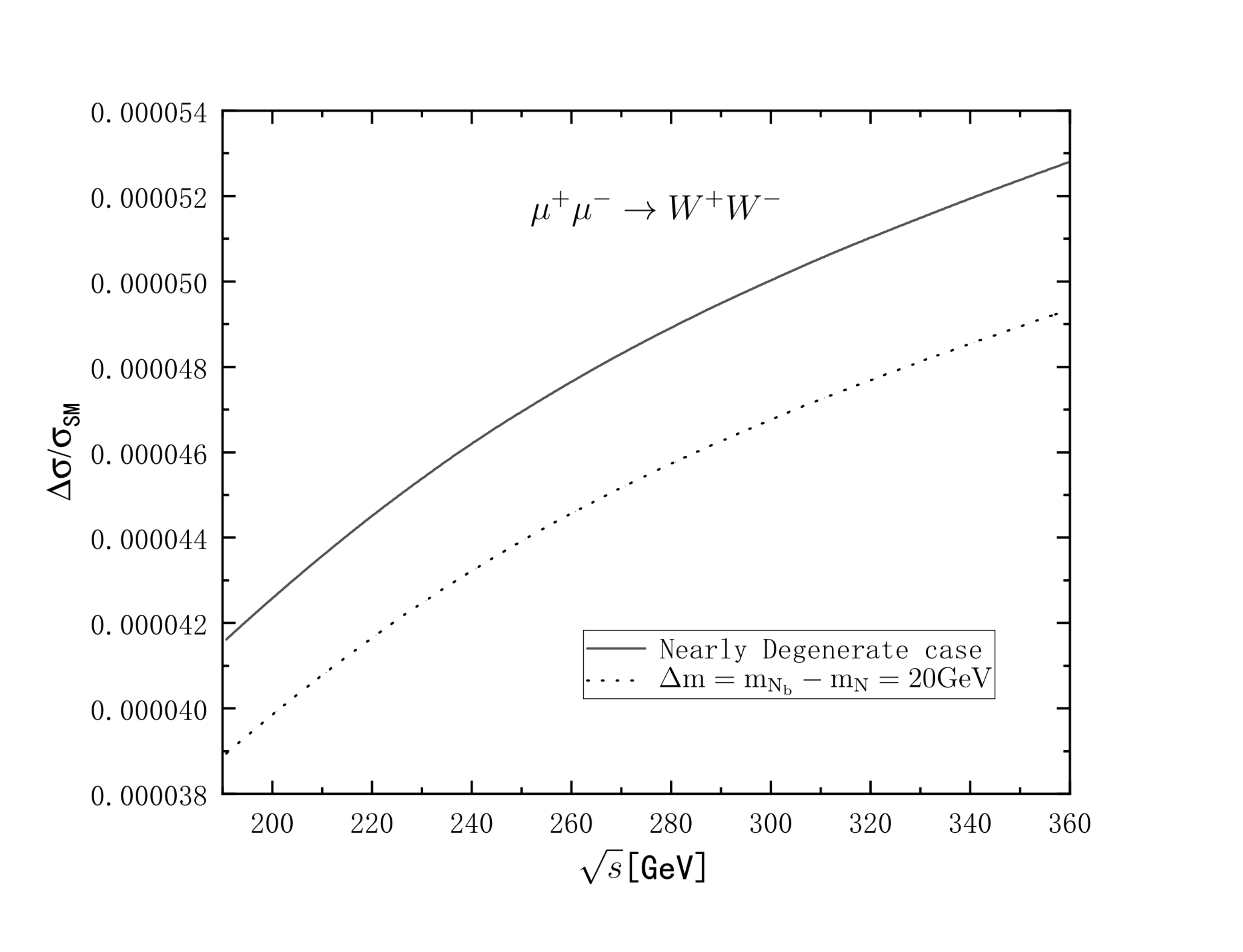} }
		\caption{ The cross section $\Delta\sigma=\sigma_{total}-\sigma_{SM}$ of (a) $e^+e^-\rightarrow W^+W^-$ and (b) $\mu^+\mu^-\rightarrow W^+W^-$ compared with the cross section of SM part $\Delta\sigma/\sigma_{SM}$ as the functions of centre-of-mass energy $\sqrt{s}$. Black curves represent the case that two heavy Majorana neutrinos are nearly degenerate, and spot lines curves correspond to the case that $\Delta m=m_{N_b}-m_N=20$ GeV where $m_N=15$ GeV.}\label{fig1}
	\end{center}
\end{figure}

As mentioned before, many processes generating heavy Majorana neutrinos have been investigated, but their generating processes have many backgrounds that are not negligible, and some of them have effects on the CP violation, so we choice a way clean enough to generate heavy Majorana neutrinos at future lepton colliders. We find processes $e^+e^-(\mu^+\mu^-)\rightarrow W^+W^-,W^\pm\rightarrow N\ell^\pm $ are so clean that no obvious backgrounds and no CP violation in the $W$ pair poduction. For convenience, we take $\ell^\pm=\mu^\pm$ in our calculation. Firstly, we want to see if it is possible to search heavy Majorana neutrinos in the processes of $W^+W^-$ pair production. The $W$ pair production can be written as $e^+(\mu^+)(l_2)+e^-(\mu^-)(l_1)\rightarrow W^+(k_p)+W^-(k_m)$, where $l_1$, $l_2$, $k_p$ and $k_m$ are the four-momentum of the corresponding particles, and its cross section can be written as 
\begin{equation}
	\label{10}
\sigma=\frac{1}{2s}\int d{\cal L}_{ips2}\overline{\left|{\cal M}\right|^2}
\end{equation}
Here, $s$ means the  center-of-mass energy squared and $d{\cal L}_{ips2}$ represents the two-body Lorentz invariant phase space of the final particles. $\overline{\left|{\cal M}\right|^2}$ is the squared scattering amplitude averaged (summed) over the initial (final) particles, which can be expressed as		
\begin{align}
	\label{11}
	 \overline{\left|{\cal M}\right|^2} = \; & \frac{1}{4}(\overline{\left|{\cal M}_{SM}\right|^2}+\overline{\left|{\cal M}_{NP}\right|^2}+2Re\overline{\left[{\cal M}_{SM} {\cal M}_{NP}^\ast\right]})\; ,\\
\label{12}
	 \overline{\left|{\cal M}_{SM}\right|^2} =\; & g^4\biggl\{\text{sin}^4\theta_w\left|D_\gamma(q)\right|^2{\cal T}_\gamma+\left|D_Z(q)\right|^2{\cal T}_Z+\frac{1}{4}\left|D_\nu(p_n)\right|^2\nn \\
	 &-\text{sin}^2\theta_w{\cal F}_{12}+\frac{1}{2}\text{sin}^2\theta_w{\cal F}_{13}-\frac{1}{2}{\cal F}_{23}
	\biggr\}\; , \\
	\label{13}
	 \overline{\left|{\cal M}_{NP}\right|^2} =\; & \frac{g^4\left|R_{eN}\right|^4}{4}\biggl\{{\cal T}_n\left(\left|D_N(p_n)\right|^2+\left|D_{N_b}(p_n)\right|^2\right)+{\cal F}_{45}\biggr\}\; , \\
	 \label{14}
	 2Re\overline{\left[{\cal M}_{SM} {\cal M}_{NP}^\ast\right] }=\; & g^4\biggl\{\frac{\text{sin}^2\theta_w\left|R_{eN}\right|^2}{2}\left({\cal F}_{14}+{\cal F}_{15}\right)-\frac{\left|R_{eN}\right|^2}{2}\left({\cal F}_{24}+{\cal F}_{25}\right)\nn \\
	 &+\frac{\left|R_{eN}\right|^2}{4}{\cal T}_n\left({\cal F}_{34}+{\cal F}_{35}\right)
	 \biggr\}\; , 	  
\end{align}
where $q=l_1+l_2$, $p_n=l_1-k_m$, $\overline{\left|{\cal M_{SM}}\right|^2}$ is the contribution from SM part, $\overline{\left|{\cal M_{NP}}\right|^2}$ is the contribution from New Physics (NP) part and $\overline{2Re\left[{\cal M}_{SM} {\cal M}_{NP}^\ast\right]}$ is the part from the interference between the contribution of SM part and NP part. The explicit expressions of $D_\gamma$, $D_Z$, $D_\nu$, $D_N$, $D_{N_b}$, and ${\cal T}_\gamma$, ${\cal T}_Z$, ${\cal T}_n$ and ${\cal F}_{12}$, ${\cal F}_{13}$, ${\cal F}_{23}$, ${\cal F}_{45}$, ${\cal F}_{14}$, ${\cal F}_{15}$, ${\cal F}_{24}$, ${\cal F}_{25}$, ${\cal F}_{34}$ and ${\cal F}_{35}$ in the amplitude are shown in Appendix~\ref{appA}.
In order to see obvious CP violation, we take two heavy Majorana neutrinos in nearly degenerate case at most of time except that when we study the influence of $\Delta m$ on CP violation and on the cross section of $e^+e^-(\mu^+\mu^-)\rightarrow W^+ W^-$ in NP part.
The $W^\pm$ pairs are produced directly from this kind of lepton-antilepton collision. In the SM way of lepton collision, there are no heavy Majorana neutrinos in the decay channels, in order to consider the contribution of physics beyond SM, we take two right-handed Majorana neutrinos into consideration as propagator of the t channel in lepton-antilepton collision. We calculate the cross section of $W$ pair production, and the result of $\Delta\sigma/\sigma_{SM}$ is shown in Fig.~\ref{fig1}, where $\Delta\sigma=\sigma_{total}-\sigma_{SM}$, $m_N=15$ GeV.
 The black lines represent that two neutrinos are nearly degenerate, and the spot lines correspond to
$\Delta m=m_{N_b}-m_N=20$ GeV where they are not nearly degenerate. (a) and (b) represent the difference between cross section in $e^+e^-\rightarrow W^+W^-$ and $\mu^+\mu^-\rightarrow W^+W^-$. We can see that the $\Delta\sigma$ represents the contribution of NP part which is too small when compared to the cross section of SM part, so it is hard to study Majorana neutrinos in $W^+W^-$ pair production sector, we should search them at $W$ rare decay. 

\begin{figure}[!h]
	\begin{center}
		\subfigure[]{\label{fig2a}
			\includegraphics[width=0.47\textwidth]{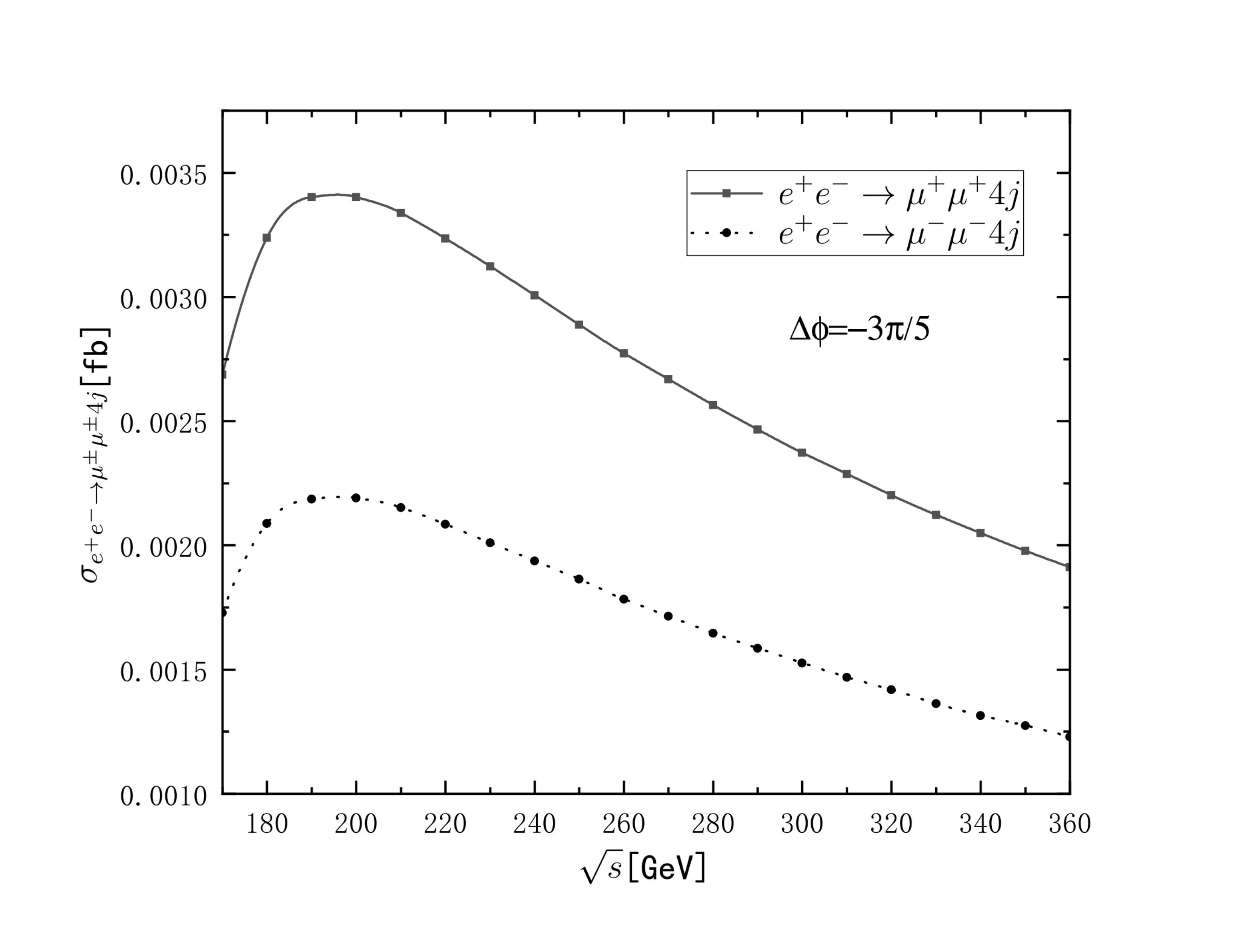} }
		\hspace{-0.5cm}~
		\subfigure[]{\label{fig2b}
			\includegraphics[width=0.47\textwidth]{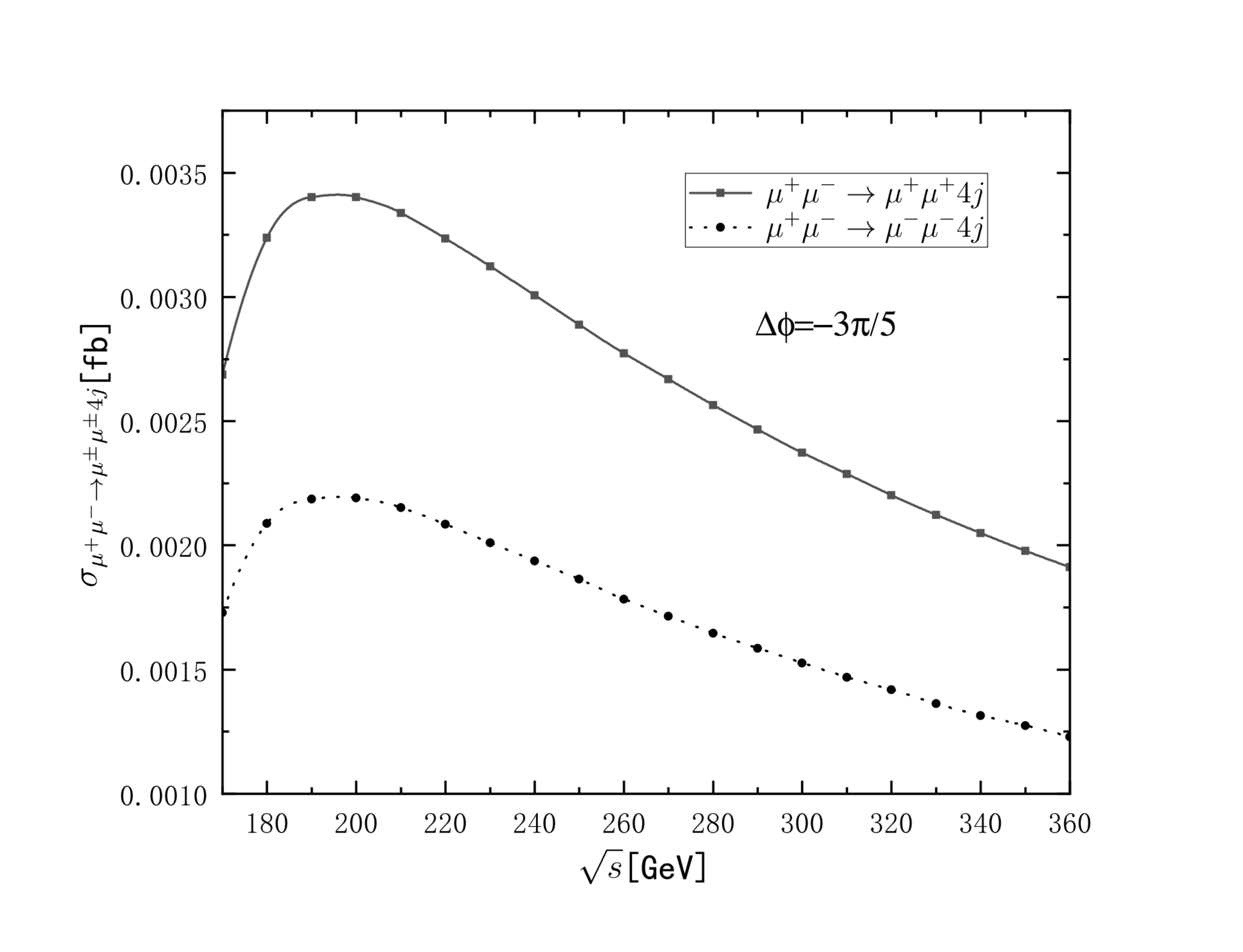} }
		\caption{Total cross section of (a) $e^+e^-\rightarrow \mu^\pm\mu^\pm 4j$ and (b) $\mu^+\mu^-\rightarrow \mu^\pm\mu^\pm 4j$ as functions of $\sqrt{s}$, CP phase is set to $\Delta\phi=\phi_{\ell N_b}-\phi_{\ell N}=-3\pi/5$ and $m_N=15$ GeV, where two heavy Majorana neutrinos are nearly degenerate.}\label{fig2}
	\end{center}
\end{figure}

\begin{figure}[!h]
	\begin{center}
		\subfigure[]{\label{fig3a}
			\includegraphics[width=0.47\textwidth]{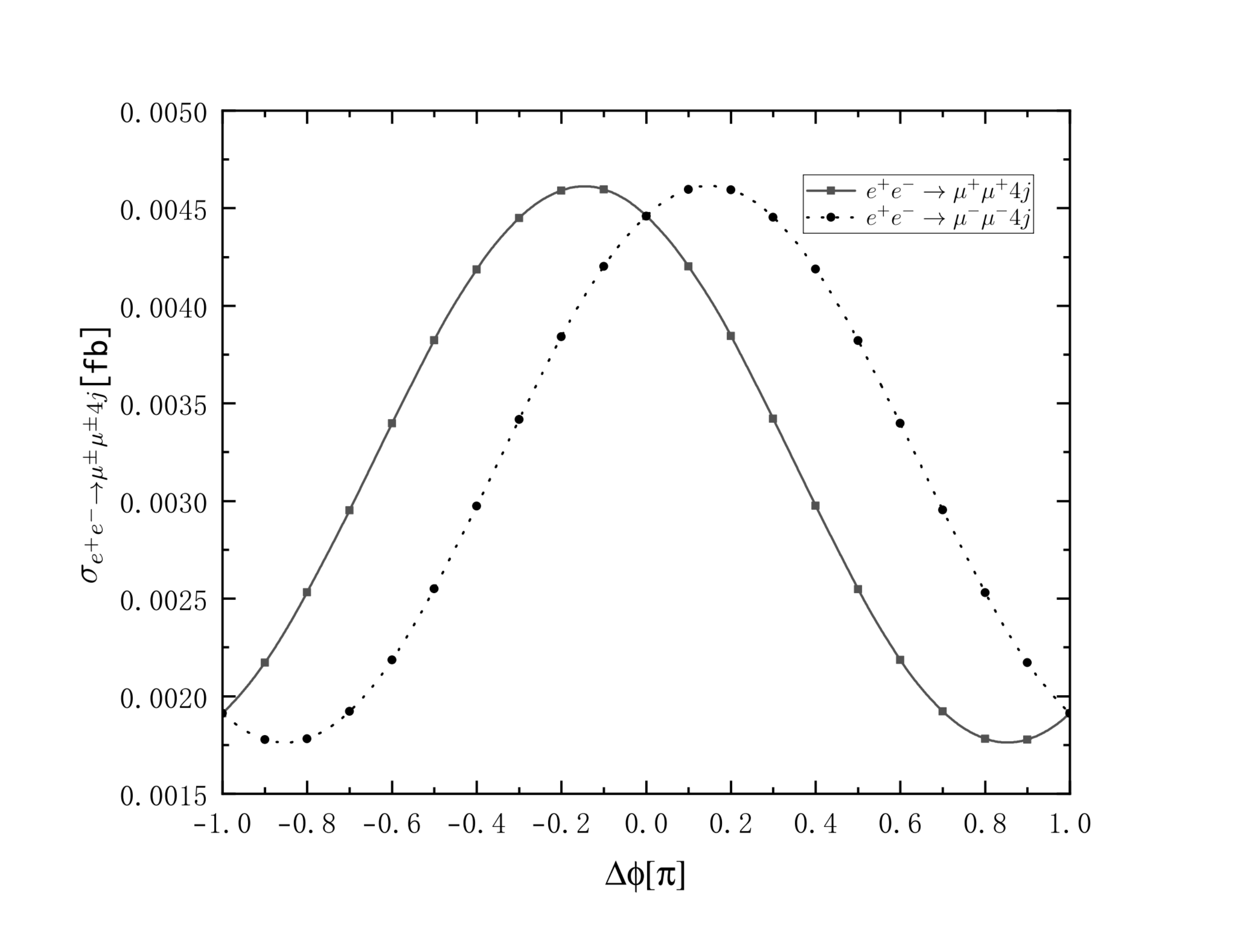} }
		\hspace{-0.5cm}~
		\subfigure[]{\label{fig3b}
			\includegraphics[width=0.47\textwidth]{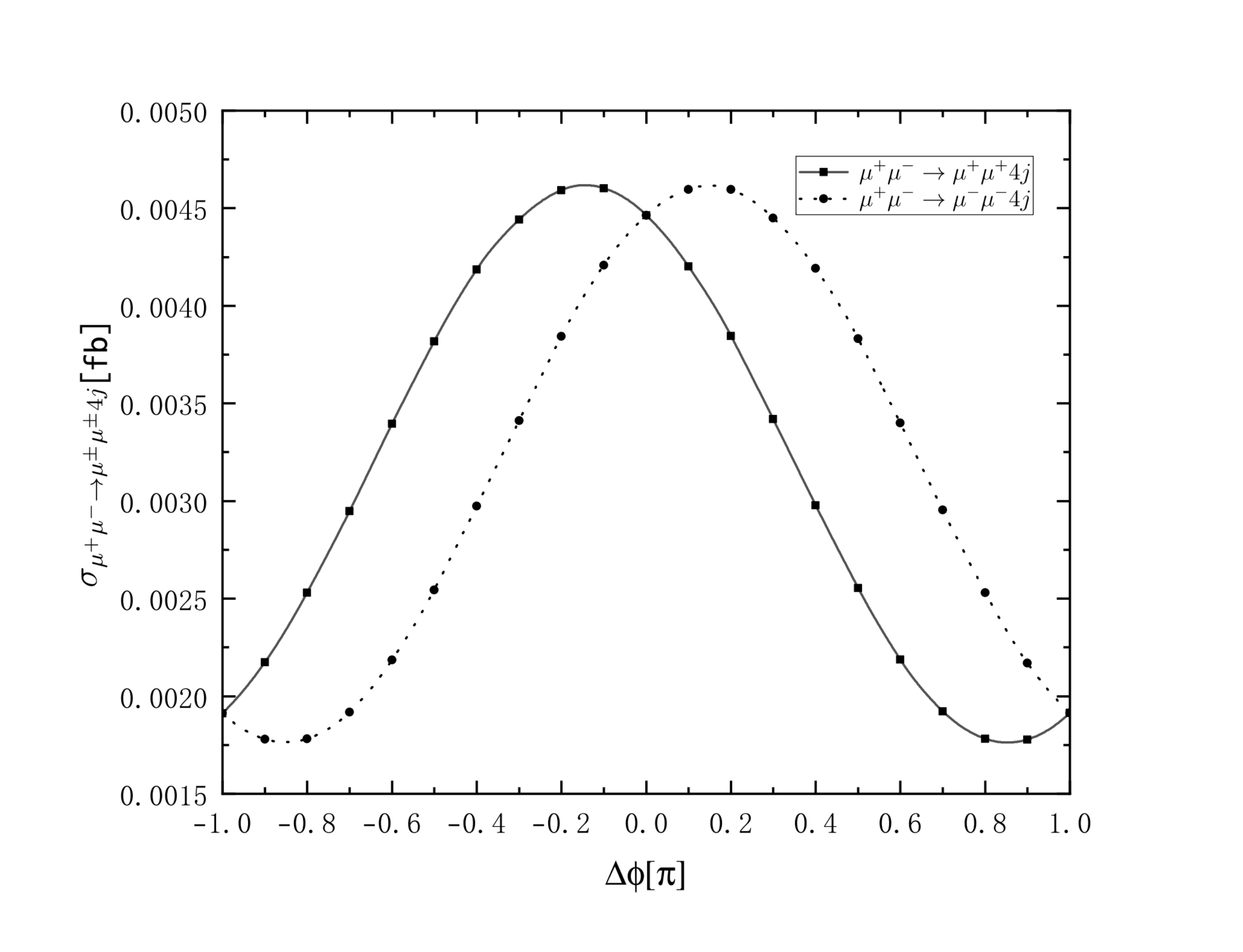} }
		\caption{Total cross section of (a) $e^+e^-\rightarrow \mu^\pm\mu^\pm 4j$ and (b) $\mu^+\mu^-\rightarrow \mu^\pm\mu^\pm 4j$ as functions of CP phase $\Delta\phi$ and $\sqrt{s}=190$ GeV, $m_N=$15 GeV, where two heavy Majorana neutrinos are nearly degenerate.}\label{fig3}
	\end{center}
\end{figure}

\begin{figure}[!h]
	\begin{center}
		\subfigure[]{\label{fig4a}
			\includegraphics[width=0.47\textwidth]{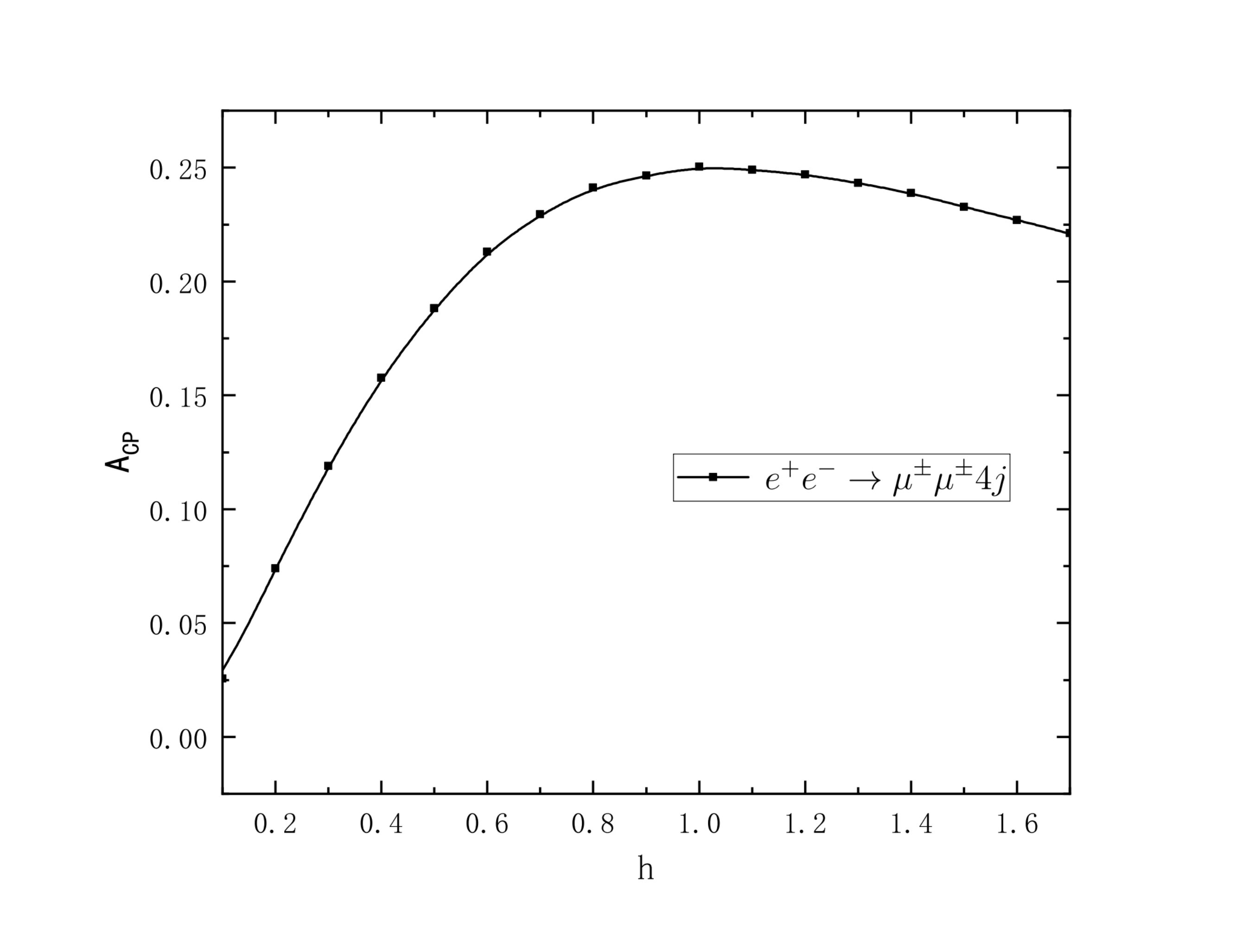} }
		\hspace{-0.5cm}~
		\subfigure[]{\label{fig4b}
			\includegraphics[width=0.47\textwidth]{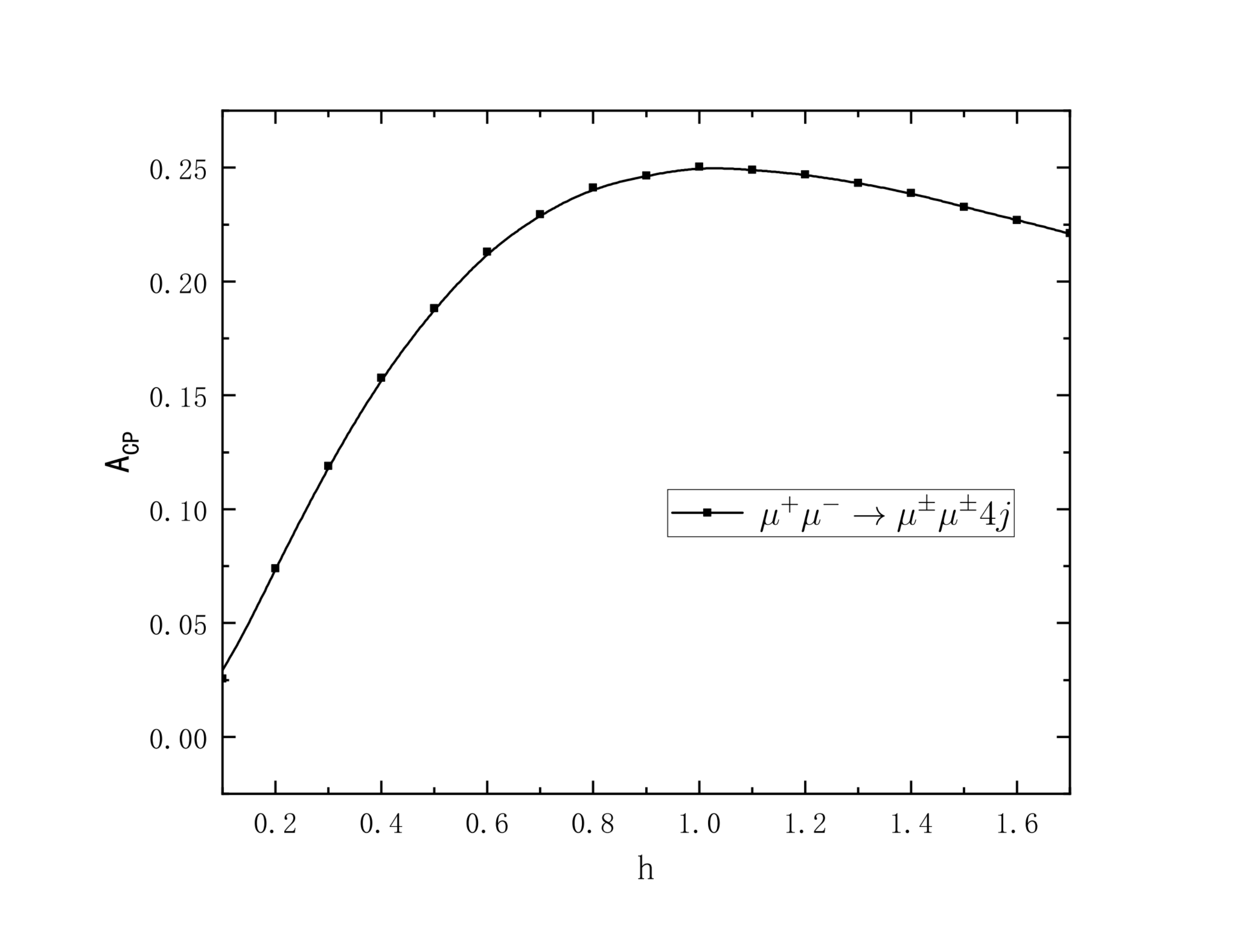} }
		\caption{$\text{A}_{\text{CP}}$ of (a) $e^+e^-\rightarrow \mu^\pm\mu^\pm 4j$ and (b) $\mu^+\mu^-\rightarrow \mu^\pm\mu^\pm 4j$ as function of $\Delta m=m_{N_b}-m_N$, $\Delta\phi=-\pi/2$, $m_N=15$ GeV, $\sqrt{s}=190$ GeV.}\label{fig4}
	\end{center}
\end{figure}

We calculated the total cross section of the process $e^+e^-(\mu^+\mu^-)\rightarrow W^\pm W^\mp\rightarrow\ell^\pm\ell^\pm+4j$. In this work we take $\ell = \mu$, and the situation of $\ell = e$ can be analysed by the same method. 
According to Eq.~\ref{7}, there are phases in mixing parameters $R_{\ell k}$.  When there are interferences between different Majorana heavy neutrinos, there will appear one term we call CP phase $\Delta\phi=\phi_{\ell N_b}-\phi_{\ell N}$. This phase will also influences the total cross section and it will cause the difference between cross section of $e^+e^-\rightarrow\ell^+\ell^+4j$ and $e^+e^-(\mu^+\mu^-)\rightarrow\ell^-\ell^-4j$ which leads to the CP violation. The results of the total cross section of $e^+e^-\rightarrow \ell^\pm\ell^\pm 4j$ and $\mu^+\mu^-\rightarrow \ell^\pm\ell^\pm 4j$ are shown in Fig.~\ref{fig2},
where the total cross sections are functions of $\sqrt{s}$. In order to see the difference between $e^+e^-(\mu^+\mu^-)\rightarrow \ell^+\ell^++4j$ and $e^+e^-(\mu^+\mu^-)\rightarrow \ell^-\ell^-+4j$ obviously, the CP phase is set to $\Delta\phi=-3\pi/5$ where the CP violation is close to the maximum value \cite{Lu:2022pvw}. The cross sections reach the maxmium about $\sqrt{s}=190$ GeV. In order to see the influence of CP phase $\Delta\phi$ on cross section, we put the result of cross section as a function of $\Delta\phi$ in Fig.~\ref{fig3}. 
We can see that the total cross section of $e^+e^-(\mu^+\mu^-)\rightarrow W^+W^-\rightarrow \mu^+\mu^+4j$ and $e^+e^-(\mu^+\mu^-)\rightarrow W^+W^-\rightarrow\mu^-\mu^-4j$ reach the maximum at about $\Delta\phi=-0.1\pi$ and $\Delta\phi=0.1\pi$ respectively. We also calculate the events and total cross section at future lepton colliders for $e^+e^-(\mu^+\mu^-)\rightarrow\ell^+\ell^+4j$ and $e^+e^-(\mu^+\mu^-)\rightarrow\ell^-\ell^-4j$ with $\Delta\phi=-\pi/10$ and $\Delta\phi= \pi/10$  separately at $\sqrt{s}=$ 190GeV and $m_{N_1}=15$ GeV. The events for $e^+e^-\rightarrow\ell^\pm\ell^\pm 4j$ with the luminosity of ${\cal L}=3000 fb^{-1}$ can reach 13 with $\sigma_{total}=4.6\times10^{-3}fb$, and for $\mu^+\mu^-\rightarrow\ell^\pm\ell^\pm 4j$ with the luminosity of ${\cal L}=1000 fb^{-1}$ can reach 4 with $\sigma_{total}=4.6\times10^{-3}fb$. So it is possible to find the heavy Majorana neutrinos at future lepton colliders at the luminosity of ${\cal L}=1000 fb^{-1}$ or higher.
Then we analyse the CP violation in the process $\ell^+\ell^-\rightarrow\mu^+\mu^+4j$, where $\ell^+\ell^-=e^+e^-$ or $\mu^+\mu^-$, and we define the CP violation as 
\begin{eqnarray}
	\label{15}
	{\cal A}_{CP}=\frac{Br(\ell^+\ell^-\rightarrow \mu^+\mu^+4j)-Br(\ell^+\ell^-\rightarrow \mu^-\mu^-4j)}{Br(\ell^+\ell^-\rightarrow \mu^+\mu^+4j)+Br(\ell^+\ell^-\rightarrow \mu^-\mu^-4j)}\;,
\end{eqnarray}
then we define $h=\Delta m/\Gamma_{N_a}$, where the $\Gamma_{Na}$ is the total decay width of heavy Majorana neutrino $N_a$ and its influence on CP violation is put in Fig.~\ref{fig4}, we can see that CP violation of $e^+e^-\rightarrow \mu^\pm\mu^\pm 4j$ reaches the maximum value at about $h=1$, and it's the same in $\mu^+\mu^-\rightarrow \mu^\pm\mu^\pm 4j$.

\section{SUMMARY}\label{sec4}

The neutrino oscillation experiments show that neutrinos have minor non-zero masses, it insinuates the existence of new physics beyond the SM. If we extend the SM by introducing the heavy Majorana neutrinos and allowing for lepton number violation, we can explain the tiny neutrino masses via seesaw mechanisms and the baryon asymmetry of the Universe. In this paper, we investigate the prospects for searching for heavy Majorana neutrinos in rare $W$ decays at future lepton colliders, such as CLIC, ILC and MuC. It is a LNV process which can be a probe to find new physics. There exists CP violation in this kind of rare decays of $W$ pair. The significant interferences of contributions from two different Majorana neutrinos produce the CP violation between $W^+\rightarrow\ell^+_\alpha\ell^+_\beta(\bar{q}q^\prime)^-$ and $W^-\rightarrow\ell^-_\alpha\ell^-_\beta(q\bar{q}^\prime)^+$. In this work we analyse the process of $W$ pair production by $e^+e^-$ and $\mu^+\mu^-$ collisions at CLIC and MuC, where we add two heavy Majorana neutrinos as the contributions of new physics part. The cross section of NP part is too small in $W$ pair production and it is not possible to search heavy Majorana neutrinos here, we need to find them in $W$ rare decays. We find that at the situation that the $\Delta m$ is considerable, the cross section contributed by NP part is much smaller when they are nearly degenerate. We find that the total cross section of process goes to the maxium value at about $\sqrt{s}=190$ GeV. It is found that at $\sqrt{s}=190$ GeV there will be 13 events with the luminosity of ${\cal L}=3000 fb^{-1}$ for $e^+e^-\rightarrow\ell^+\ell^+4j$ and 4 events with the luminosity of ${\cal L}=1000 fb^{-1}$ for $\mu^+\mu^-\rightarrow\ell^+\ell^+4j$. The investigation in this paper maybe helpful to search for the new physics signal with respect to the LNV process via the heavy Majorana neutrinos at future high energy lepton colliders.

\section*{ACKNOWLEDGEMENTS}
The authors thank the member of the Institute of theoretical physics of Shandong University for their helpful discussions. This work is supported in part by National Natural Science Foundation of China (Grants No.12235008, 11875179) and Natural Science Foundation of Shandong Province(ZR2021QA040).
\begin{appendix}
	
	\section{Calculation of the squared scattering amplitude }\label{appA}

In this appendix, we show explicitly the squared scattering amplitude given in Eq.~\ref{11}-Eq.~\ref{14}.The functions ${\cal F}_{12}$, ${\cal F}_{13}$,${\cal F}_{23}$, ${\cal F}_{45}$, ${\cal F}_{14}$, ${\cal F}_{15}$,${\cal F}_{24}$, ${\cal F}_{25}$, ${\cal F}_{34}$ and ${\cal F}_{35}$ can be respectively expressed as

\begin{align}
	\label{A1}
	&{\cal F}_{12}={\cal T}_{\gamma Z}\left(D_\gamma D_Z^\star+D_ZD_\gamma^\star\right)\; , \\
	\label{A2}
	&{\cal F}_{13}={\cal T}_{\gamma n}\left(D_\gamma D_\nu^\star+D_\nu D_\gamma^\star\right)\; , \\
	\label{A3}
	&{\cal F}_{14}={\cal T}_{\gamma n}\left(D_\gamma D_N^\star+D_ND_\gamma^\star \right)\; , \\
	\label{A4}
	&{\cal F}_{15}={\cal T}_{\gamma n}\left(D_\gamma D_{N_b}^\star+D_{N_b}D_\gamma^\star \right)\; , \\
	\label{A5}
    &{\cal F}_{23}={\cal T}_{Zn}\left(D_Z D_{\nu}^\star+D_{\nu}D_Z^\star \right)\; , \\
    \label{A6}
    &{\cal F}_{24}={\cal T}_{Zn}\left(D_Z D_{N}^\star+D_{N}D_Z^\star \right)\; , \\
    \label{A7}
    &{\cal F}_{25}={\cal T}_{Zn}\left(D_Z D_{N_b}^\star+D_{N_b}D_Z^\star \right)\; , \\
    \label{A8}
    &{\cal F}_{34}={\cal T}_{n}\left(D_\nu D_{N}^\star+D_{N}D_\nu^\star \right)\; , \\
    \label{A9}
    &{\cal F}_{35}={\cal T}_{n}\left(D_\nu D_{N_b}^\star+D_{N_b}D_\nu^\star \right)\; , \\
    \label{A10}
    &{\cal F}_{45}={\cal T}_{n}\left(D_N D_{N_b}^\star+D_{N_b}D_N^\star \right)\; , 
\end{align}
where the $D_X(p^2)$ is the Breit-Wigner propagator and can be defined as

\begin{align}
	\label{A11}
	D_X(p^2)=\frac{1}{p^2-m^2_X+im_X\Gamma_X}\; ,
\end{align}
with $m_X$ and $\Gamma_X$ being the mass and total decay width of the corresponding particles.

The explicit expression of ${\cal T}_\gamma$, ${\cal T}_Z$, ${\cal T}_n$, ${\cal T}_{\gamma Z}$, ${\cal T}_{\gamma n}$, ${\cal T}_{Z n}$ can be given by

\begin{align}
	\label{A13}
	{\cal T}_{\gamma}= & -24m_W^2s-\frac{2s^2t(s+t)}{m_W^4}-24(s^2+st+t^2)+\frac{2s(3s^2+4st+t^2)}{m_W^2}\;, \\
	\label{A14}
	{\cal T}_{Z}=& \frac{(4sin^2\theta_w-8sin^4\theta_w-1)}{4m_W^4}\biggl\{12m_W^6s+s^2t(s+t)+12m_W^4(s^2+st+t^2)-\nn \\
	&m_W^2s(3s^2+4st+4t^2)\biggr\}\;, \\
	\label{A15}
	{\cal T}_{n}=&-m_W^2+(s-5t)t+\frac{(s-2t)t^2}{m_W^2}-\frac{t^3(s+t)}{m_W^4}\;,\\
	\label{A16}
	{\cal T}_{\gamma Z}= & \frac{(1-4sin^2\theta_w)}{2m_W^4}\biggl\{12m_W^6s+s^2t(s+t)+12m_W^4(s^2+st+t^2)-m_W^2s(3s^2+4st+4t^2)\biggr\}\;,\\
	\label{A17}
	{\cal T}_{\gamma n}= & -6m_W^2s+\frac{st^2(s+t)}{m_W^4}-3(s^2+2t^2)-\frac{t(2s^2+st+2t^2)}{m_W^2} \;,\\
	\label{A18}
	{\cal T}_{Zn}= & \frac{1-2sin^2\theta_w}{2m_W^4}\biggl\{6m_W^6s-st^2(s+t)+3m_W^4(s^2+2t^2)+m_W^2t(2s^2+st+2t^2)\biggr\}\;,
\end{align}
where $s=2(l_1\cdot l_2)$, $t=-2(l_2\cdot k_p)$.

\end{appendix}

\end{document}